\documentclass[a4paper,twoside]{article}
\usepackage[authoryear]{natbib}
\bibpunct{(}{)}{;}{a}{}{,}
\usepackage{graphicx}
\date{} 
\def\aj{AJ}%
\def\apj{ApJ}%
\def\apjs{ApJS}%
\def\aap{A\&A}%
\def\mnras{MNRAS}%
\title{\large\bf\flushleft When Darwin Met Einstein: Gravitational Lens Inversion with Genetic Algorithms}
\author{\parbox{\textwidth}{\flushleft
\vspace{-0.5cm}
{\it Brendon J. Brewer \& Geraint F. Lewis}\\
\vspace{0.45cm}
{\small Institute of Astronomy, School of Physics, University of Sydney, NSW 2006, Australia}\\
{\small Email: {\tt brewer,gfl@physics.usyd.edu.au}}}}
\begin{document}
\twocolumn[
\begin{changemargin}{.8cm}{.5cm}
\begin{minipage}{.9\textwidth}
\vspace{-1cm}
\maketitle
%
%
\small{\bf  Abstract:  Gravitational  lensing  can magnify  a  distant
source,  revealing structural detail  which is  normally unresolvable.
Recovering  this  detail through  an  inversion  of  the influence  of
gravitational lensing, however, requires optimisation of not only lens
parameters,  but also of  the surface  brightness distribution  of the
source.   This  paper  outlines  a  new approach  to  this  inversion,
utilising genetic  algorithms to  reconstruct the source  profile.  In
this  initial   study,  the  effects  of  image   degradation  due  to
instrumental and  atmospheric effects are neglected and  it is assumed
that the  lens model  is accurately known,  but the  genetic algorithm
approach   can  be   incorporated  into   more   general  optimisation
techniques,  allowing the optimisation  of both  the parameters  for a
lensing model and the surface brightness of the source.}


\medskip
\medskip
\end{minipage}
\end{changemargin}
] \small

\section{Introduction}

Gravitational lensing is an  important astrophysical tool, mapping the
distribution  of  matter  on   many  scales  and  revealing  typically
unresolved  detail  of  distant  sources  through  magnification  [see
\citet{newk}  for a review  of lensing  physics]. Early
studies concentrated upon the properties of multiply imaged quasars to
determine the  underlying mass distribution within  the lensing galaxy.
However, the point-like nature of quasars (which results in point-like
images) present  only a  limited amount of  constraints on  the lensing
mass  distribution, with  a  number of  degenerate  solutions able  to
explain             the             observed             configuration
\citep{1988AJ.....96.1570K,1988AJ.....95.1619S,1991ApJ...373..354K}.

If the  source in a  gravitational lensing is extended,  the resulting
image  is  also  extended  and  each  resolution  element  effectively
provides  a  constraint on  any  modelling.   Any modelling,  however,
becomes more complex than the  simple case of point-like lensing which
asks  the question  ``what distribution  of  mass in  the lensing  can
account for the observed  image locations and brightnesses?''. With an
extended  source,   the  question  has  to  be   rephrased  as  ``what
distribution of  mass in the  lensing galaxy {\it and  distribution of
brightness  in  the  source}   can  account  for  the  observed  image
configuration?''. To  answer this  question, more novel  approaches to
gravitational     lensing    modelling     have     been    undertaken
\citep{1989MNRAS.238...43K,  1992ApJ...401..461K, 1996ApJ...464..556E,
1996ApJ...465...64W,  2003ApJ...590..673W, 2004MNRAS.349....1W, wayth,
dye}.   These  techniques  typically   determine  not  only  the  mass
distribution in  the lensing galaxy,  but also the  surface brightness
distribution in the source.

Given  the  form  of  the  lens  and  the  source,  it  is  relatively
straightforward to compute  the resultant image by using  a simple ray
tracing method. The inverse problem, which is what occurs in practice,
is much  more difficult to  solve.  Furthermore, inverse  problems are
fraught with the  question of solution uniqueness. To  this end, it is
advisable to  tackle a  particular problem with  a range  of inversion
techniques  and  compare  the  various  outcomes;  if  all  approaches
converge to  the same result, some  faith can be given  to the overall
solution.  Currently,  the repertoire of  gravitational lens inversion
techniques  is  relatively  small   and  so  this  paper  presents  an
alternative  approach  to   gravitational  lens  modelling,  utilising
genetic   algorithms  to   reconstruct  the   source.    This  initial
investigation  of  this approach,  the  effects  of image  degradation
through  instrumental  and  atmospheric  effects  (i.e.   seeing)  are
neglected,  although these  can  be implemented  in a  straightforward
fashion.     The    approach     is    described    in    detail    in
Section~\ref{approach},     considering    a     perfect,    noiseless
image. Section  ~\ref{results} discusses the influence  of the various
parameters  influencing  the  inversion  technique,  also  considering
reconstruction of noisy images. Section~\ref{model} considers the more
general  problem  of  the  optimization  of both  the  source  surface
brightness   distribution  and  the   parameters  governing   the  mass
distribution  in  the  lensing  galaxy.  This paper  closes  with  the
conclusions which are presented in Section~\ref{conclusions}.

\begin{figure}
\begin{center}
\includegraphics[scale=0.7, angle=0]{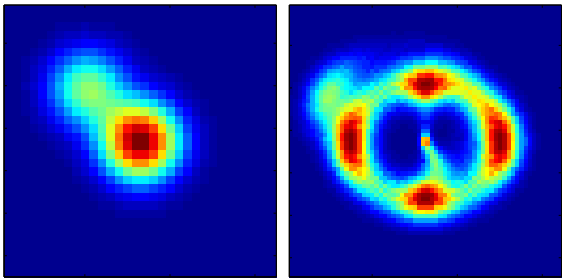}
\caption{The left-hand  panel presents the  artificial source utilised
in this study,  while the right-hand panel presents  the image of this
source   as   seen  through   the   model   gravitational  lens   (see
section~\ref{lensmodel}).   Note  that these  images  as on  different
scales, with the source panel being half the width of the image panel.
The source  is defined on a  grid of 32$\times$32  pixels, whereas the
resultant  image is 64$\times$64  pixels.  For  observed gravitational
lens  images,  the  image  would  be $\sim3$  arcseconds  on  a  side,
corresponding to  a pixel scale  of $0.05$ arcseconds per  pixel.  The
corresponding source  plane region possesses  an image scale  of 0.025
arcseconds   per  pixel  and   a  side   length  of   0.8  arcseconds.
}\label{fig1}
\end{center}
\end{figure}

\section{Approach}\label{approach}

\subsection{Genetic Algorithms}\label{gas}
Genetic algorithms  are an approach  to problems of  optimisation that
take their inspiration from evolutionary biology [for a popular review
of genetic  algorithms and other aspects of  biological computing, see
\citet{levy}].   The  basic  approach,  presented in  some  detail  in
\citet{1995ApJS..101..309C}, mimics the evolutionary struggle of life,
with different  individuals having different  probabilities of passing
on  their  genes  to  the  next  generation,  with  the  probabilities
dependent on  the environment and the physical  characteristics of the
organism (which in turn depend on  the genes). Humans have long used a
basic knowledge of  heredity to breed for desirable  traits in animals
and crops, even  before Charles Darwin proposed his  theory of natural
selection \citep{species}. Genetic algorithms take this idea and apply
it by “breeding” better solutions to the problem at hand.

In terms of  the algorithmic  approach, several  features are
required;
\begin{itemize}
\item {\bf Encoding:} Each potential  solution to a problem is encoded
into a {\it genome}. This is  often represented as a series of digits,
but can be a simple bit string.
\item  {\bf Expression:}  This  decodes the  information  in the  {\it
genome} into  a {\it phenotype}.  For many applications,  this decodes
the  {\it genome}  into a  series  of real  numbers that  are used  as
parameters for a particular model.
\item {\bf  Fitness evaluation:} This compares the  {\it phenotype} of
the {\it genome}  to the problem, assigning a  quantitative measure of
the  goodness of  fit  of  the potential  solution  (e.g.  a  standard
$\chi^2$ measurement).
\end{itemize}
With these, an initial population of genomes, each
representing  a potential  solution to  the  problem at hand, can  be
generated. Typically this involves assigning each genome with a random
sequence of  digits or bits. In  evolving this population  to the next
generation, several steps are involved:

\noindent
{\bf  a)  Ranking  \&  Selection  Pressure:} The  goal  of  a  genetic
algorithm  is  to produce  subsequent  generations  of solutions  with
greater  fitness by  ensuring that  the  fittest member  of a  current
population to pass their genetic material onto to the next generation.
As  noted by \citet{1995ApJS..101..309C},  as evolution  proceeds, the
average  and maximum  fitness of  a population  continually increases.
The spread in fitness, however, decreases, with the overall population
becoming  homogeneous.  With  such  uniform fitness  in a  population,
simple selection  on fitness  alone effectively samples  randomly from
the population  and evolution  stalls.  To circumvent  this, selection
must be  made relative  to the current  population.  To this  end, the
population is ranked in terms of its fitness, with the least fit being
assigned  a value  of 0,  whereas the  most fit  possesses a  value of
1. Members  are   selected  from   this  current  generation   with  a
probability dependent upon their ranking, such that
\begin{equation}
p \propto (ranking)^\beta
\label{selection}
\end{equation}
where  $\beta$ is called  the selection  pressure.  If  $\beta=0$, the
probability for selection is  uniform throughout the population, while
larger  values  of  $\beta$  preferentially select  only  the  fittest
members in the population.

\noindent
{\bf  b) Elitism:}  The fittest  member of  the current  generation is
cloned and represents  the first member of the  next generation. This
ensures that  the maximum fitness of subsequent  generations can never
fall.
  
\noindent
{\bf c) Breeding:} Further members of the next generation are produced
by breeding  the members of  the current generation, with  the genetic
information of the current population used to determine the genomes of
the next. The probability that  an individual will breed is based upon
its ranking  and the selection  pressure as outlined  previously.  Two
breeding  strategies  are  adopted,  {\it asexual}  and  {\it  sexual}
reproduction. With  {\it asexual}  reproduction, a selected  member of
the current generation is  cloned, preserving the genetic information,
to  provide a  new member  of the  next generation.   In  {\it sexual}
reproduction, two members of the current generation are selected and a
new  individual  is  formed  with  the combination  of  their  genetic
material; a random portion of the genome of one parent is ``copied and
pasted'' over the  corresponding part of the other  parent's genome to
produce  the  resulting  offspring  genome; this  allows  the  genetic
material of  successful organisms to  mix. Whether the creation  of an
offspring is due to sexual reproduction is determined randomly, with a
pre-defined probability referred to as the {\it crossover rate}.

\noindent
{\bf  d) Mutation:}  A  population which  reproduces purely  asexually
rapidly becomes dominated by a single genome and evolution grinds to a
halt.  Random  mutations in the  genetic sequence can  drive evolution
beyond this  point, increasing diversity  in a population~\footnote{It
should be  noted that the neo-Darwinistic  view that it  is solely DNA
that evolves has been  questioned, with the implication that evolution
is  actually  a  complex  interplay  of  the  genotype  and  phenotype
\citep[see][]{collapse}.} The  probability that a  particular digit in
the  genetic   sequence  is  mutated  (straight   after  breeding)  is
determined by the {\it mutation rate}.

\begin{figure}
\begin{center}
\includegraphics[scale=0.9, angle=0]{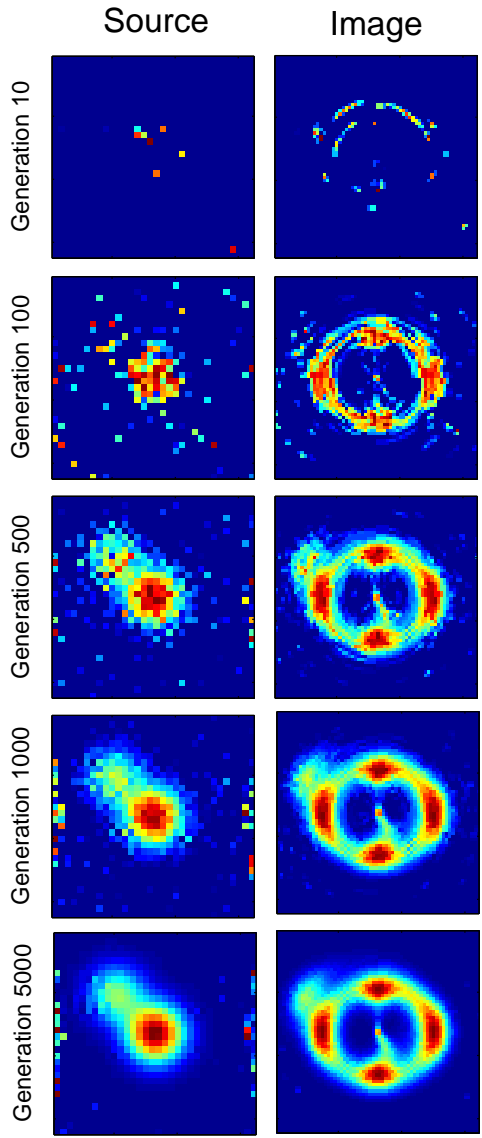}
\caption{An  example of  the evolutionary  sequence obtained  with the
genetic  algorithm reconstruction of  an idealised  gravitational lens
system.   The  left-hand  panel  displays  the  source  plane  surface
brightness distribution  at 10, 100,  500, 1000 and  5000 generations,
while  the  right-hand  panel   presents  the  resulting  image  plane
distribution.  It is clear,  when comparing to Figure~\ref{fig1}, that
the fitness of  the solution is increasing with  each generation. Note
that the vertical noise at the  edge of the source plane correspond to
regions  which are  not mapped  into  the image  plane and  so do  not
contribute     to    the    overall     fitness    of     the    image
reconstruction.}\label{evolve}
\end{center}
\end{figure}

The  breeding and  mutation are  continued until  a new  generation is
formed, and the entire initial  generation is culled. Steps a) through
d) are  then repeated,  with subsequent generations  exhibiting fitter
solutions  to  the  problem.   The  evolution is  terminated  when  an
appropriate fitness criterion is satisfied.

\citet{1995ApJS..101..309C}  and \citet{1995A&A...296..164H} presented
some of the earliest  applications of genetic algorithms in astronomy,
with the former  detailing a freeware algorithm (PIKAIA)~\footnote{\tt
http://www.hao.ucar.edu/public/research/si/pikaia}.     These    early
studies focused upon the fitting of light curves and galactic rotation
curves \citep{1995ApJS..101..309C},  and constraining accretion stream
mapping  in eclipsing  polars  \citep{1995A&A...296..164H}, but  since
then the application of genetic algorithms has been used in the design
of           filters            and           filter           systems
\citep{1998MNRAS.299..176O,2004A&A...419..385B},     modelling     the
structure of  the galaxy \citep{2003AJ....125.1958},  the signature of
gamma  ray   bursts  \citep{2001MNRAS.328..951P},  solar  oscillations
\citep{2003MNRAS.346..825F}  and  even  the scheduling  of  telescopes
\citep{2003A&A...403..357G}.

\subsection{Gravitational Lens Inversion}\label{inversion}

\subsubsection{Encoding \& Phenotyping}
For the purposes of this  project, each individual genome was taken to
be a string of 1024  characters.  The expressed genome (the phenotype)
represented 32$\times$32  pixels, where each pixel could  take a value
between  0 and  $F_{max}$.  This  pixel array  represents  the surface
brightness distribution  of the source.  Such an  encoding ensures the
surface brightness is subject to  a positivity constraint [this is not
the    case    of     some    other    inversion    approaches    e.g.
\citet{2003ApJ...590..673W}].     In     the    coming    simulations,
$F_{max}=255$.

\begin{figure}[h]
\begin{center}
\includegraphics[scale=0.45, angle=0]{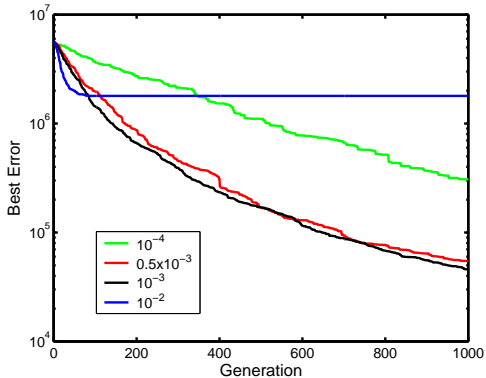}
\caption{The influence of  the mutation rate on fitness  as a function
of  generation. The  box in  this  panel (and  in subsequent  figures)
denotes the different values  of parameter adopted.  Note, this figure
presents  $fitness^{-1}$  as  defined  in  Equation~\ref{fitness}.   A
reasonable rate  of mutation  is needed to  stimulate progress  in the
population  by  providing  random  variations  for  selection  to  act
upon. However, if the mutation  rate is too high, any ``good'' genomes
will be severely  affected by too many mutations.   The black curve is
for the adopted mutation rate of $10^{-3}$.}\label{fig3}
\end{center}
\end{figure}

\subsubsection{Gravitational Lens Model}\label{lensmodel}
The scaled lensing equations relating  a position $(x,y)$ in the image
plane to the corresponding point  $(x_s,y_s)$ in the source plane [see
\citet{1998LRR.....1...12W}] can  be written  in terms of  a potential
$\phi(x,y)$ (related to the mass distribution of the lens) such that

\begin{equation}
  \begin{array}{rcl}
	x_s = x - \frac{\partial \phi}{\partial x}|_{(x,y)}\\
	\\	
	y_s = y - \frac{\partial \phi}{\partial y}|_{(x,y)}
  \end{array}
\label{interp}
\end{equation}

The gravitational lens potential was  chosen to be the three parameter
pseudo-isothermal  elliptic  potential \citep{1989MNRAS.238...43K}  of
the form
\begin{equation}
\phi(x,y) = b \sqrt{ r_c^2 + (1-\epsilon)x^2 + (1+\epsilon)y^2 }
\end{equation}
where $r_c$  is the  core radius of  the potential, $\epsilon$  is the
ellipticity  and $b$  is  an overall  normalisation factor  (typically
linked to the  velocity dispersion of the lensing  galaxy).  While the
determination of these parameters is the typical goal of many analyses
of  gravitational  lenses,  in  this  intial  examination  of  genetic
algorithms, it is  assumed that the model is  fixed and its parameters
are known.   In other words, the  goal is to find  the source profile,
given the observed image and the  form of the lens. The adopted values
for the  lens parameters were $b=0.5$,  $\epsilon=0.25$ and $r_c=0.1$.
The  full  optimisation  problem  of  the  source  profile  and  model
parameters will be discussed in Section~\ref{model}.

\subsubsection{Fitness Determination}
To  test the  effectiveness  of the  genetic  algorithmic approach  to
gravitational lens  inversion, an  example solution was  defined. This
consisted of two offset Gaussian  profiles of differing heights and is
displayed  graphically in  the left-hand  panel  of Figure~\ref{fig1},
while the image  of this source as seen  through the lensing potential
outlined previously appears in the  right-hand panel. The image of the
source is realised upon a 64$\times$64 grid.

\begin{figure}
\begin{center}
\includegraphics[scale=0.45, angle=0]{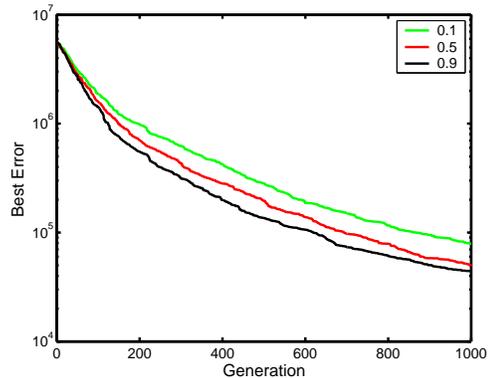}
\caption{As  for  Figure~\ref{fig3}, but  detailing  the influence  of
cross-over rate on fitness as a function of generation.}\label{fig4}
\end{center}
\end{figure}

In  determining the  fitness  of a  particular  genome, its  expressed
phenotype is used to represent a potential source. This is mapped, via
the lens model, to produce  the resulting image configuration. This is
then  compared to the  {\it ideal}  image (Figure~\ref{fig1})  and the
assigned fitness  was chosen to  be the reciprocal  of the sum  of the
squared differences  between the reconstructed image  and the observed
image:
\begin{equation}
fitness = \frac{1}{\sum_{i=1}^{64} \sum_{j=1}^{64} ( m_{ij} - p_{ij} )^2}
\label{fitness}
\end{equation}
where $m_{ij}$ are the pixel  brightnesses of the model generated from
the genome and $p_{ij}$ are the pixel brightnesses of the ``observed''
image.   Clearly, in  the absence  of noise,  the sum  of  the squared
differences  for a perfect  genome is  zero and  hence the  fitness is
infinite. Of course,  for a real gravitational lens  system, this {\it
ideal} image will be replaced with a noisy observed image. Simulations
of the genetic  algorithm considering the influence of  noise, will be
discussed in Section~\ref{noise}.

\section{Results}\label{results}

Examples  of  the source  reconstructions  achieved  with the  genetic
algorithm are presented in Figure~\ref{evolve}. These were produced by
optimising the  source profile, assuming  that the correct  lens model
was known. From top to bottom, each pair of panels presents a snapshot
of  the evolution,  showing the  fittess member  of a  generation. The
left-hand  panel graphically  presents the  surface brightness  of the
source  (the  genome),  whereas  the  right-hand  panel  presents  the
resultant  image  configuration  (the  phenotype).   Clearly,  as  the
population  is  evolved to  older  generations,  the  accuracy of  the
solution  increases (compare  the  source and  image configuration  at
Generation 5000  to Figure~\ref{fig1}).  Note that  the `noisy' pixels
along   the   vertical  sides   of   the   source  reconstruction   in
Figure~\ref{evolve} correspond  to regions  which are not  mapped into
the image region and so does  not contribute to the overall fitness of
the solution.

\subsection{Evolutionary Parameters}

There  are several  parameters that  can affect  the performance  of a
genetic algorithm.  It is important  to find a set of parameters which
is  good   at  finding  a  solution   quickly,  since  it   can  be  a
computationally expensive  task to run  for many generations.  Also it
may be necessary to evolve the population many times, in which case it
is very important to minimise the time required to find a satisfactory
solution.  There  are several  factors  which  have  the potential  to
influence this  performance, and the four which  are investigated here
are the mutation rate, the  crossover rate, the size of the population
and  the selection  pressure, which  is the  relationship  between the
fitness score and the probability of being selected for breeding.

It is known that there  are interactions between these parameters such
that, for  example, the  answer to the  question ``what  mutation rate
should  I  use?''  depends  on  the values  of  the  other  parameters
\citep{1995ApJS..101..309C}. It is also highly dependent on the nature
of the problem  at hand. Therefore, the best that can  be hoped for is
some  general picture  of  what order  of  magnitude these  parameters
should be.

\begin{figure}
\begin{center}
\includegraphics[scale=0.45, angle=0]{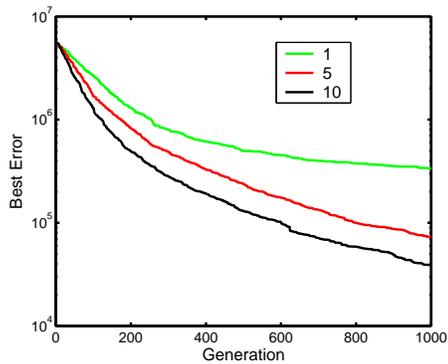}
\caption{As  for  Figure~\ref{fig3}, but  detailing  the influence  of
selection    pressure   on    the   fitness    as   a    function   of
generation.}\label{fig5}
\end{center}
\end{figure}

To  test the  dependency of  the  genetic alorithmic  approach on  the
adopted parameters, an initial  reconstruction was undertaken with the
following  parameters;  the  crossover   rate  was  set  to  0.9,  the
population size was 50, the  mutation rate was set to $10^{-3}$
and the  selection pressure  was set to  10 (these were  ‘good’ values
found  after much  painstaking  trial and  error).   In the  following
sections, one  of these parameters  was varied, while the  others were
kept  fixed, allowing  the influence  of  the varied  parameter to  be
determined.

\begin{figure}
\begin{center}
\includegraphics[scale=0.45, angle=0]{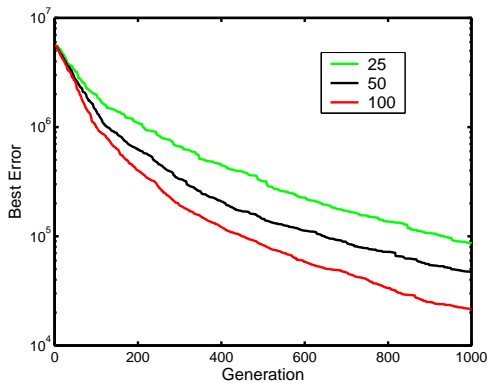}
\caption{As  for  Figure~\ref{fig3}, but  detailing  the influence  of
population    size    on    the    fitness   as    a    function    of
generation.}\label{fig6}
\end{center}
\end{figure}

\subsubsection{Mutation}

Figure~\ref{fig3}  presents  the   evolution  of  the  ``Best  Error''
(defined as $fitness^{-1}$) of the fittest member of a population as a
function of generation for several differing mutation rates.  Clearly,
if  the mutation  rate  is  low $(10^{-4})$  then  the genome  evolves
slowly, but shows steady  improvement. Increasing the rate of mutation
increases the rate of improvement  in the genome. However, this cannot
continue  indefinitely;  as  can   be  seen,  a  large  mutation  rate
$(10^{-2})$  results in  a  rapid increase  in improvement  initially,
after  a short time  the evolution  stagnates. This  is because  for a
mutation rate  of 0.01 and a  genome length of 1024,  on average there
are about 10 mutations per  generation on each individual, and this is
enough  to  outweigh  any   improvements  that  have  evolved  through
selection.  Hence, a little  mutation is  a good  thing, but  too much
mutation is not. A general rule  of thumb is that a good mutation rate
is one which will give about 1 mutation over the whole genome.

\subsubsection{Cross-over Rate}

Figure~\ref{fig4} presents the results  of varying the cross-over rate
on the rate  of improvement of the genome. For  the three trial values
of 0.1,  0.5 and 0.9, there is  very little difference on  the rate of
improvement over time. Clearly, the algorithm is generally insensitive
to the adopted value of the cross-over rate.

\subsubsection{Selection Pressure}

Figure~\ref{fig5}  presents the influence  of the  selection pressure,
$\beta$,  on the  rate of  improvement of  the fittest  genome.  Three
$\beta$ values  of 1, 5 and  10 were trialed;  remember that $\beta=0$
ensures a uniform selection probability for breeding from a population
(i.e.   no  selection   pressure  at   all),  whereas   larger  values
preferentially  selected the  fittest  members for  breeding. For  the
adopted  values, there  was a  definite advantage  in using  a $\beta$
value  bigger than  1, but  only  a slight  difference in  performance
between $\beta$ = 5 and $\beta$ = 10.

\begin{figure*}
\begin{center}
\includegraphics[scale=0.7, angle=0]{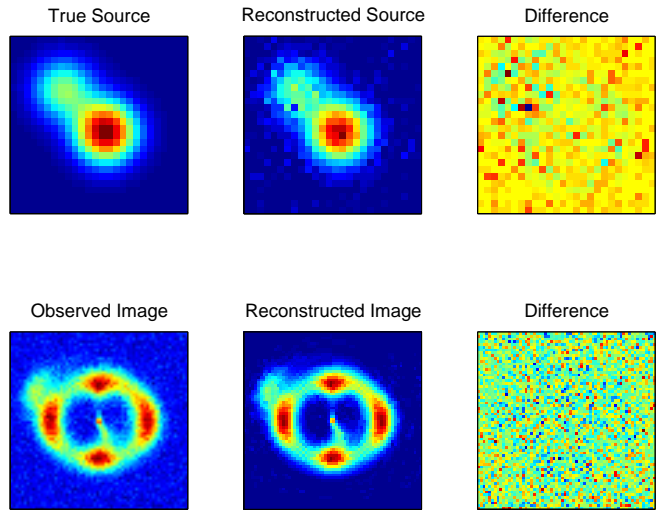}
\caption{The results of the  genetic algorith reconstruction using the
noisy image after 5000 generations.}\label{noisy}
\end{center}
\end{figure*}

\subsubsection{Population Size}

The number  of individuals in the  population can also  be varied, and
the   influence  of   changing   population  size   is  presented   in
Figure~\ref{fig6}. Clearly larger populations  have a larger spread in
genetic variation and it is seen that the larger populations do evolve
more rapidly.   From a computational  point of view,  however, smaller
populations  result  in  a  significant  speed  advantage,  with  less
calculations required per generation.

In fact, as the time taken for each generation is roughly proportional
to the population size, the  larger genetic variability seen in larger
population  can be  outweighing  the time  required  to calculate  the
fitness of a population.

\subsection{Source Reconstruction \& Noise}\label{noise}

In real  life, astronomical images  are always contaminated  with some
random  noise, arising from  sources such  as thermal  fluctuations in
electronics, the  effects of the atmosphere, and  even photon counting
noise if the source is faint (a  feature that is shared by a number of
extended gravitational lenses). Therefore it is worthwhile to test how
the genetic algorithm performs when the observed image is contaminated
with noise.

Including  ‘errors’ into  the observed  image also  sets  a meaningful
criterion  to  decide whether  a  particular  reconstruction is  “good
enough”. This criterion will be met if the reconstructed image matches
the observed  image to within the  statistical error level  set by the
noise.   For this test,  a normally  distributed random  variable with
$\sigma$ = 5  was added to each pixel in  the observed image, creating
the   noisy   image   shown   in   the  lower   left-hand   panel   of
Figure~\ref{noisy}.

\begin{figure}
\begin{center}
\includegraphics[scale=0.45, angle=0]{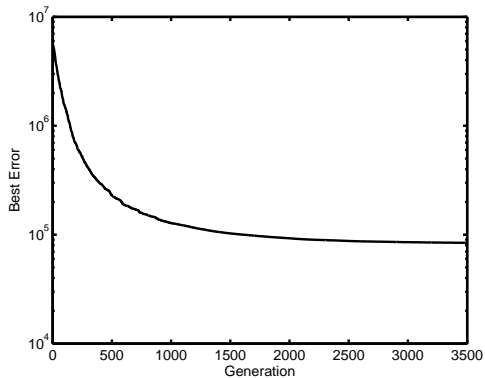}
\caption{Error  of the fittest  solution vs  generation for  the noisy
image.  Notice  that   it  does  not  tend  to   zero,  as  a  perfect
reconstruction of this image is impossible.}\label{noisy2}
\end{center}
\end{figure}

When  the  genetic  algorithm  was  run,  the  error  vs.   time  plot
(Figure~\ref{noisy2}) flattened  out at a much higher  value. This was
not surprising,  because in  this case  it is not  possible to  have a
source  that reproduces the  observed image  exactly, with  its random
fluctuations  between neighbouring  pixels. The  additional  panels in
Figure~\ref{noisy}  presents   the  reconstructed  image   and  source
profile, as well  as the difference between these  and the true source
and  corresponding image; these  are consistent  with the  input noise
characteristics.

This  process  can be  thought  of as  a  curve-fitting  problem in  2
dimensions, with $64^2$ data points (the observed image), each with an
error bar of 5  units. The aim is to fit this  data using a model that
has $32^2-N$ free parameters (each pixel of the source; note $N\sim80$
corresponds to  those pixels in the  source grid which  are not lensed
into  the final image,  and hence  are not  true free  parameters; see
Figure~\ref{evolve}).   The chi-squared statistic  for this  fit (with
degrees of freedom equal to the number of constraints minus the number
of free paramaters in the model, i.e. $64^2 - 32^2 + 80$) is given by

\begin{equation}\label{stats}
\chi^2  =  \sum_{i=1}^{64}  \sum_{j=1}^{64}  \frac{( m_{ij}  -  p_{ij}
)^2}{\sigma^2} = \frac{1}{\sigma^2*fitness}
\end{equation}

Hence, a  reconstruction is  statistically good (within  1$\sigma$) if
the sum of the squared  differences between the observed image and the
image  of the  reconstructed source  is within  $\sigma^2  \times (\nu
\pm\sqrt{2 \nu})$,  where $\nu$ is  the number of degrees  of freedom.
\citep{1992nrfa.book.....P} Hence,  a good fit corresponds  to a value
in  the range  of $78800\pm1985$.  The genetic  algorithm was  able to
reduce the error to $\sim82000$, and so recovered an acceptable fit to
the noisy image.

\section{Full Optimisation}\label{model}

\subsection{Lens Parameters in the Genome}
In  general,  the goal  of  gravitational  lens  reconstruction is  to
determine both the surface brightness of the source and the parameters
describing the mass distribution in the deflecting galaxy. How can the
genetic algorithmic  approach be  generalised to tackle  this problem?
One idea that was tried  was incorporating the lens parameters as part
of the  potential solutions  to the problem,  by encoding them  in the
genomes. Then, the evolution  would hopefully select individuals whose
lens parameters  were close to the  right values, and  then proceed to
optimise the source pixels.  The  result of this approach was not very
successful.  What  actually occurred was  that early on,  a particular
value  of  parameters was  “locked  on”  and  became dominant  in  the
population.  Then,  the sources  were optimised for  those (incorrect)
values, and there was little hope in ever getting to the right values.
This was  because any change to  the parameter values  would require a
huge chance jump in the sources in order to gain a higher fitness than
what had already evolved.

\subsection{Direct Search Method}
While incorporating the lens parameters in the genome was found not to
be successful, an approach  using a started direct search, independent
of the genetic algorithm, was found to be successful. For the purposes
of this study, simple one-dimensional searches were employed, although
the  technique  can  be  easily generalised  into  higher  dimensional
searches.

\begin{figure}
\begin{center}
\includegraphics[scale=0.45, angle=0]{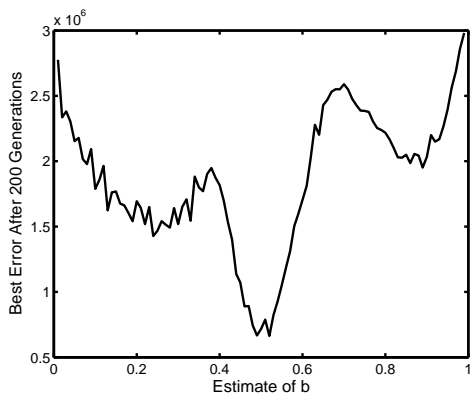}
\caption{The best  error after 3500  generations as a function  of the
value  of $b$. Clearly,  a good  reconstruction of  the image  is only
possible  if  the  value  of  $b$  is  close  to  the  true  value  of
0.5.}\label{direct}
\end{center}
\end{figure}

An  example   of  such  a  one-dimensional  search   is  presented  in
Figure~\ref{direct}, in which $r_c$  and $\epsilon$ were held at their
optimum  value, and a  suite of  reconstructions were  undertaken with
differing  values of  $b$. This  figure presents  the residual  of the
fittest  solution  of the  population  after  3500 generations.   This
clearly reveals that a good  reconstruction is only possible if $b$ is
very close to the correct value of  0.5.

\begin{figure}
\begin{center}
\includegraphics[scale=0.45, angle=0]{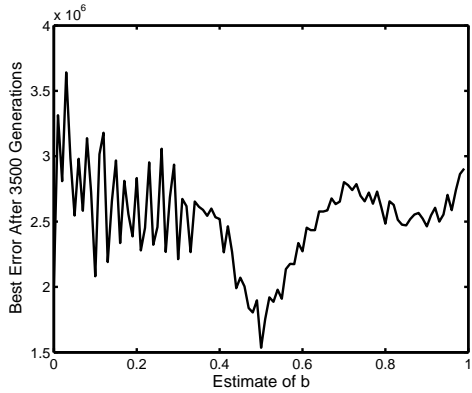}
\caption{Best error after  200 generations as a function  of the value
of $b$. Note  that the evolution is proceeding at  a faster rate, even
at    this    early    stage,    if   the    lens    parameters    are
correct.}\label{direct2}
\end{center}
\end{figure}

The  major   disadvantage  of  this  approach   is  its  computational
inefficiency. Each  point in Figure~\ref{direct}  required the genetic
algorithm  to be  run  over  3500 generations,  which  takes about  10
minutes on  a modern  desktop computer. However,  it was  noticed that
even after a small number of generations, the correct value of $b$ was
``winning the race'' and so a smaller number of generations can be run
to  determine the  interesting  regions of  parameters  space for  for
further   exploration.   To   illustrate   this,  Figure~\ref{direct2}
presents  the  same  result  as Figure~\ref{direct},  using  only  200
generations  rather  than  3500.  This  obviously  has  a  significant
computational advantage.

\begin{figure}
\begin{center}
\includegraphics[scale=0.45, angle=0]{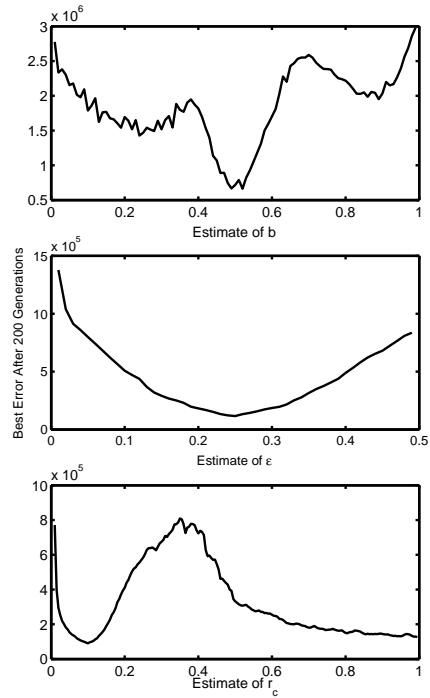}
\caption{Best  error  after  200  generations  as a  function  of  the
estimated values  of the parameters (with  the other two  fixed at the
correct values). The minima of the plots are all at the true values of
the parameters.  These figures  were calculated without  resetting the
population after each 200 generations.}\label{smooth}
\end{center}
\end{figure}

Calculating  each  point  in  Figures~\ref{direct}  and  \ref{direct2}
required  one run  of the  genetic algorithm,  and the  population was
reset each time  such that the initial genomes  are sequences of zeros
(corresponding  to black  pixels).  However,  if each  point  in these
figures are calculated sequentially, the populations need not be reset
as the  solution to  the previous point  already contains  fairly good
solution,  but for  a slightly  different $b$-value.   Then the  GA is
given a  “head start” in  trying to improve the  solution, effectively
tweaking  the  previous solution  to  produce  a  new solution.   This
results    in   a    dramatic   smoothing    out   of    the   figures
(Figure~\ref{smooth}),  and  allows  the  parameter to  be  calculated
correctly  to and  accuracy about  3 significant  figures in  only 200
generations.

As  noted earlier,  it should  be possible  to simply  generalise this
one-dimensional method to include more than one free parameter, either
by using a large grid search  (which may take a long time, although it
is straightforward to  devise a parallel computing scheme  to do this,
since each run can be done independently of the others), or by using a
multidimensional minimisation method such as Powell's method.

\section{Conclusions}\label{conclusions}
This  paper  has introduced  a  new  technique  for the  inversion  of
gravitational lensed images of extended sources. This utilises genetic
algorithms to evolve an  optimal source for a particular gravitational
lens model.  It  is seen that this approach  successfully recovers the
source    configuration   of    an   idealised    gravitational   lens
system. Furthermore, it was demonstrated that this genetic algorithmic
approach successfully  recovers the source profile in  the presence of
noise  and can be  incorporated into  more general  gravitational lens
optimisation schemes.  This  initial investigation has considered only
a simple  model of gravitational lensing,  neglecting detailed aspects
of true gravitational  lens systems, such as various  sources of noise
and  image  smearing  due  to instrumental  and  atmospheric  effects.
However, due  to the  forward mapping of  this approach, these  can be
added in a straight  forward fashion, providing an inversion technique
that can  be applied to  observed gravitational lens systems.   Due to
the limited time-frame  of this initial project, these  aspects of the
algorithm will be left as further work.

The most  time consuming part  of the genetic algorithmic  approach is
the calculation of the fitness of each member in a generation, scaling
with the size of the population. For a particular genome, however, the
calculation of the fitness is  independent of the other members of the
generation. This  leads to a  simple parallelisation of  the approach,
with    the   fitness   calculation    farmed   out    to   individual
processors. Furthermore,  genetic evolution  can be driven  harder via
the    inclusion   of   parasitic    organisms   or    `black   sheep'
\citep{2000A&A...357.1170B}, speeding  up the evolution  of the genome
to fitter solutions and  preventing evolutionary stagnation; these too
will be incorporated into fuller version of this inversion technique.

\section{Acknowledgements}
Aspects  of this  research were  undertaken  as a  third year  special
project at the University of Sydney.


\end{document}